\documentclass{article}
\usepackage{amsfonts}
\usepackage{lscape}
\usepackage{amsmath}
\usepackage{amssymb}
\usepackage{graphicx}

\begin{document}

\begin{titlepage}
\vspace*{3cm}
\begin{center}
{\Large \textsf{\textbf{New treatment of the noncommutative Dirac equation with a Coulomb potential}}}
\end{center}
\vskip 5mm
\begin{center}
{\large \textsf{Lamine Khodja$^{a}$ and Slimane Zaim$^{b}$}}\\
\vskip 5mm
$^{a}$Laboratoire de Physique Th\'{e}orique, D\'{e}partement de Physique,\\
Universit\'{e} A. Mira de Béjaia, Route Targa Ouzemour, Béjaia, Algeria.\\
$^{b}$ D\'{e}partement des Siences de la Mati\'{e}re, Facult\'{e} des Sciences,\\
Universit\'{e} Hadj Lakhdar -- Batna, Algeria.\\
\end{center}
\vskip 2mm
\begin{center}{\large\textsf{\textbf{Abstract}}}\end{center}
\begin{quote}
Using the approach of the modified Euler-Lagrange field equation together with the corresponding Seiberg-Witten maps of the dynamical fields, a noncommutative Dirac equation with a Coulomb potential is derived. We then find the noncommutative modification to the energy levels and the possible new  transitions. In the nonrelativistic limit a general form of the hamiltonian of the hydrogen atom is obtained, and we show that the noncommmutativy plays the role of spin and magnetic field which gives the hyperfine structure.
\end{quote}
\vspace*{2cm}
\noindent\textbf{\sc Pacs numbers}: 11.10.Nx, 81Q05, 03.65.Pm, 31.30.jx
\end{titlepage}

\section{Introduction}

The connection between string theory and the noncommutativity \cite{1, 2, 3,
4} motivated a large amount of work to study and understand many physical
phenomenon. There is a flurry of activity in analysing divergences \cite{5},
unitarity violation \cite{6}, causality \cite{7}, and new physics at very
short distances of the Planck-length order \cite{8}.

The noncommutative field theory is characterised by the commutation relations
between the noncommutative coordinates themselves; namely:
\begin{equation}
\left[ \hat{x}^{\mu },\hat{x}^{\nu }\right] _{\ast }=i\theta ^{\mu \nu },
\end{equation}%
where $\hat{x}^{\mu }$ are the coordinate operators and $\theta ^{\mu \nu }$
are the non-commutativity parameters of dimension of area that signify the
smallest area in space that can be probed in principle. The Groenewald-Moyal
star product of two fields $f\left( x\right) $ and $g\left( x\right) $ is
given by

\begin{equation}
f\left( x\right) \ast g\left( x\right) =\exp \left( \frac{i}{2}\frac{%
\partial }{x^{\mu }}\frac{\partial }{y^{\nu }}\right) f\left( x\right)
g\left( y\right) \mid _{y=x}
\end{equation}

The most obvious natural phenomena to use in hunting for noncommutative
effects are simple quantum mechanics systems, such as the hydrogen atom \cite%
{9,10,11}. In the noncommutative space one expects the degeneracy of the
initial spectral line to be lifted, thus one may say that non-commutativity
plays the role of spin.

In a previous work \cite{12}, by solving the deformed Klein-Gordon equation
in canonical non-commutative space, we showed that the energy is shifted:
the first term of the energy correction is proportional to the magnetic
quantum number, which behavior is similar to the Zeeman effect as applied to
a system without spin in a magnetic field; the second term is proportional
to $\theta ^{2}$, thus we explicitly accounted for spin effects in this
space.

The purpose of this paper is to study the extension of the Dirac field in
the same context by applying the result obtained to a hydrogen atom.

This paper is organized as follows. In section 2 we propose an invariant
action of the noncommutative Dirac field in the presence of an
electromagnetic field. In section 3, using the generalised Euler-Lagrange
field equation, we derive the deformed Dirac equation. In section 4, we
apply these results to the hydrogen atom, and by the use of the perturbation
theory, we solve the deformed Dirac equation and obtain the noncommutative
modification of the energy levels. In section 5, we introduce the
non-relativistic limit of the noncommutative Dirac equation and solve it
using perturbation theory and deduce that the non-relativistic noncommutative
Dirac equation is the same as the Schr\"{o}dinger equation on noncommutative
space. Finally, in section 5, we draw our conclusions.

\section{Seiberg-Witten maps}

Here we look for a mapping $\phi ^{A}\rightarrow \hat{\phi}^{A}$ and $%
\lambda \rightarrow \hat{\lambda}\left( \lambda ,A_{\mu }\right) $, where $%
\phi ^{A}=(A_{\mu },\psi )$ is a generic field, $A_{\mu }$ and $\psi $ are
the gauge field and spinor respectively (the Greek and Latin indices denote
curved and tangent space-time respectively), and $\lambda $ is the $\mathrm{U%
}(1)$ gauge Lie-valued infinitesimal transformation parameter, such that:
\begin{equation}
\hat{\phi}^{A}\left( A\right) +\hat{\delta}_{\hat{\lambda}}\hat{\phi}%
^{A}\left( A\right) =\hat{\phi}^{A}\left( A+\delta _{\lambda }A\right) ,
\label{eq:trans}
\end{equation}%
where $\delta _{\lambda }$ is the ordinary gauge transformation and $\hat{%
\delta}_{\hat{\lambda}}$ is a noncommutative gauge transformation which are
defined by:
\begin{align}
\hat{\delta}_{\hat{\lambda}}\hat{\psi}=i\hat{\lambda}\ast \hat{\psi},&
\qquad \delta _{\lambda }\psi =i\lambda \psi ,  \label{eq:tempo1} \\
\hat{\delta}_{\hat{\lambda}}\hat{A}_{\mu }=\partial _{\mu }\hat{\lambda}+i%
\left[ \hat{\lambda},\hat{A}_{\mu }\right] _{\ast },& \qquad \delta
_{\lambda }A_{\mu }=\partial _{\mu }\lambda .  \label{tempo2}
\end{align}

In accordance with the general method of gauge theories, in the
noncommutative space, using these transformations one can get at second
order in the non-commutative parameter $\theta ^{\mu \nu }$ (or equivalently
$\theta $) the following Seiberg--Witten maps \cite{1}:
\begin{align}
\hat{\psi}& =\psi +\theta \psi ^{1}+\mathcal{O}\left( \theta ^{2}\right) , \\
\hat{\lambda}& =\lambda +\theta \lambda ^{1}\left( \lambda ,A_{\mu }\right) +%
\mathcal{O}\left( \theta ^{2}\right) , \\
\hat{A}_{\xi }& =A_{\xi }+\theta A_{\xi }^{1}\left( A_{\xi }\right) +%
\mathcal{O}\left( \theta ^{2}\right) ,  \label{eq:SWM1} \\
\hat{F}_{\mu \xi }& =F_{\mu \xi }\left( A_{\xi }\right) +\theta F_{\mu \xi
}^{1}\left( A_{\xi }\right) +\mathcal{O}\left( \theta ^{2}\right) ,
\label{eq:SWM2}
\end{align}%
where
\begin{equation}
F_{\mu \nu }=\partial _{\mu }A_{\nu }-\partial _{\nu }A_{\mu }.
\end{equation}

To begin, we consider an action for a non-commutative Dirac field in the
presence of an electrodynamic gauge field in a non-commutative space-time.
We can write:
\begin{equation}
\mathcal{S}=\int d^{4}x\,\left( \overline{\hat{\psi}}\ast \left( i\gamma
^{\nu }\hat{D}_{\nu }-m\right) \ast \hat{\psi}-\frac{1}{4}\hat{F}_{\mu \nu
}\ast \hat{F}^{\mu \nu }\right) ,  \label{eq:action}
\end{equation}%
where the gauge covariant derivative is defined as: $\hat{D}_{\mu }\hat{\psi}%
=\left( \partial _{\mu }+ie\hat{A}_{\mu }\right) \ast \hat{\psi}$.

Next we use the generic-field infinitesimal transformations \eqref{eq:tempo1}
and \eqref{tempo2} and the star-product tensor relations to prove that the
action in eq. \eqref{eq:action} is invariant. By varying the scalar density
under the gauge transformation and from the generalised field equation and
the Noether theorem we obtain \cite{13}:
\begin{equation}
\frac{\partial \mathcal{L}}{\partial \hat{\psi}}-\partial _{\mu }\frac{%
\partial \mathcal{L}}{\partial \left( \partial _{\mu }\hat{\psi}\right) }%
+\partial _{\mu }\partial _{\nu }\frac{\partial \mathcal{L}}{\partial \left(
\partial _{\mu }\partial _{\nu }\hat{\psi}\right) }+\mathcal{O}\left( \theta
^{2}\right) =0.  \label{eq:field}
\end{equation}

\section{ Non-commutative Dirac equation}

In this section we study the Dirac equation for a Coulomb interaction ($-e/r$%
) in the free non-commutative space. This means that we will deal with
solutions of the U$(1)$ gauge-free non-commutative field equations \cite{14}.
For this we use the modified field equations in eq. \eqref{eq:field} and
the generic field $\hat{A}_{\mu }$ so that:
\begin{equation}
\delta \hat{A}_{\mu }=\partial _{\mu }\hat{\lambda}-ie\hat{A}_{\mu }\ast
\hat{\lambda}+ie\hat{\lambda}\ast \hat{A}_{\mu },
\end{equation}%
and the free non-commutative field equation:
\begin{equation}
\partial ^{\mu }\hat{F}_{\mu \nu }-ie\left[ \hat{A}^{\mu },\hat{F}_{\mu \nu }%
\right] _{\ast }=0,  \label{eq:freefield}
\end{equation}
where we assumed the non-commutative current to vanish everywhere in
space ($r\neq 0$) \cite{14}. Using the Seiberg-Witten maps
\eqref{eq:SWM1}--\eqref{eq:SWM2} and the choice \eqref{eq:freefield}
(static solution), we can obtain the following deformed Coulomb
potential \cite{10}:
\begin{align}
\hat{a}_{0}& =-\frac{e}{r}-\frac{e^{3}}{\,r^{4}}\theta ^{0j}x_{j}+\mathcal{O}%
\left( \theta ^{2}\right) , \\
\hat{a}_{i}& =\frac{e^{3}}{4\,r^{4}}\theta ^{ij}x_{j}+\mathcal{O}\left(
\theta ^{2}\right) .
\end{align}%
Using the modified field equations in eq. \eqref{eq:field} and the generic
field $\hat{\psi}$ so that:
\begin{equation}
\delta _{\hat{\lambda}}\hat{\psi}=i\hat{\lambda}\ast \hat{\psi},
\end{equation}%
the modified Dirac equation in a non-commutative space-time in the presence
of the vector potential $\hat{A}_{\mu }$ up to the first order of $\theta $
can be cast into:
\begin{equation}
\left( i\gamma ^{\mu }\partial _{\mu }-m\right) \hat{\psi}-e\gamma ^{\mu }%
\hat{A}_{\mu }\hat{\psi}+\frac{ie}{2}\theta ^{\rho \sigma }\gamma ^{\mu
}\partial _{\rho }\hat{A}_{\mu }\partial _{\sigma }\hat{\psi}=0.
\label{eq:KGmod}
\end{equation}

\subsection{Non-commutative space-space Dirac equation}

For a noncommutative space-space ($\theta ^{0i}=0$ where $i=1,2,3$), we do
not consider a noncommutative space-time ($\theta ^{0i}\neq 0$ ) since
several works have shown that the theory suffers lack of unitarity. See for
instance ref \cite{6}, it is easy to check that:%
\begin{align}
i\gamma ^{\mu }\partial _{\mu }-m& =i\gamma ^{0}\partial _{0}+i\gamma
^{i}\partial _{i}-m, \\
-e\gamma ^{\mu }\hat{A}_{\mu }& =+\frac{e^{2}}{r}\gamma ^{0}-\frac{e^{4}}{%
4r^{4}}\gamma ^{i}\theta ^{ij}x_{j}, \\
\frac{ie}{2}\theta ^{\rho \sigma }\gamma ^{\mu }\partial _{\rho }\hat{A}%
_{\mu }\partial _{\sigma }& =\frac{ie^{2}}{2r^{3}}\theta ^{ij}\gamma
^{0}x_{i}\partial _{j}=\frac{e^{2}}{2r^{3}}\gamma ^{0}\overrightarrow{\theta
}\cdot \overrightarrow{L}.
\end{align}%
Notice that: $\theta _{i}=\frac{1}{2}\epsilon _{ijk}\theta _{jk}$. Then the
noncommutative Dirac equation \eqref{eq:KGmod} up to $\mathcal{O}\left(
\theta ^{2}\right) $ takes the following form:
\begin{equation}
\left[ i\gamma ^{0}\partial _{0}+i\gamma ^{i}\partial _{i}-m+\frac{e^{2}}{r}%
\gamma ^{0}-\frac{e^{4}}{4r^{4}}\gamma ^{i}\theta ^{ij}x_{j}+\frac{e^{2}}{%
2r^{3}}\gamma ^{0}\overrightarrow{\theta }\cdot \overrightarrow{L}\right]
\hat{\psi}\left( t,r,\theta ,\varphi \right) =0.  \label{eq:temp1}
\end{equation}%
We can write this equation as:%
\begin{equation}
\hat{H}\hat{\psi}\left( t,r,\theta ,\varphi \right) =i\partial _{0}\hat{\psi}%
\left( t,r,\theta ,\varphi \right) .  \label{aa}
\end{equation}%
Then

\begin{equation}
\hat{H}=H_{0}+H_{pert}^{\theta },
\end{equation}%
where $H_{0}$ is the relativistic hydrogen atom hamiltonian%
\begin{equation}
H_{0}=\overrightarrow{\alpha }\left( -\overrightarrow{i\nabla }\right)
+\beta m-\frac{e^{2}}{r},  \label{a1}
\end{equation}%
and $H_{pert}^{\theta }$ is the leading-order perturbation%
\begin{equation}
H_{pert}^{\theta }=-\frac{e^{2}}{2r^{3}}\overrightarrow{\theta }\cdot
\overrightarrow{L}+\frac{e^{4}}{4}\overrightarrow{\theta }\cdot \left(
\overrightarrow{\alpha }\times \frac{\overrightarrow{r}}{r^{4}}\right) .
\label{a2}
\end{equation}

The first term of (\ref{a2}) which coincides with the one given in \cite{10}
describes the interaction spin-orbit where $\theta $ plays the role of spin.
The second term is absent in ref \cite{10} and $\theta $ here corresponds to
a magnetic field.

In the above the matrices $\overrightarrow{\alpha }$ and $\beta $ are given
by:%
\begin{equation*}
\beta =\left(
\begin{array}{cc}
I & 0 \\
0 & -I%
\end{array}%
\right) \qquad ;\qquad \alpha ^{i}=\left(
\begin{array}{cc}
0 & \sigma ^{i} \\
\sigma ^{i} & 0%
\end{array}%
\right) ,
\end{equation*}%
where $\sigma ^{i}$\ are the Pauli matrices:%
\begin{equation*}
\sigma ^{1}=\left(
\begin{array}{cc}
0 & 1 \\
1 & 0%
\end{array}%
\right) \qquad ;\qquad \sigma ^{2}=\left(
\begin{array}{cc}
0 & -i \\
i & 0%
\end{array}%
\right) \qquad ;\qquad \sigma ^{3}=\left(
\begin{array}{cc}
1 & 0 \\
0 & -1%
\end{array}%
\right) .
\end{equation*}

To investigate the modification of the energy levels by (\ref{a2}) , we use
the first-order perturbation theory. The spectrum of $H_{0}$ and the
corresponding wave functions are well known \ and given by (see \cite%
{15,16,17,18,19,20,21}):

\begin{equation}
\psi \left( r,\theta ,\varphi \right) =\left(
\begin{array}{c}
\phi \left( r,\theta ,\varphi \right) \\
\chi \left( r,\theta ,\varphi \right)%
\end{array}%
\right) =\left(
\begin{array}{c}
f\left( r\right) \Omega _{jM}\left( \theta ,\varphi \right) \\
g\left( r\right) \Omega _{jM}\left( \theta ,\varphi \right)%
\end{array}%
\right) ,
\end{equation}%
where the bi-spinors $\Omega _{jlM}\left( \theta ,\varphi \right) $ are
defined by:

\begin{equation}
\Omega _{jlM}\left( \theta ,\varphi \right) =\left(
\begin{array}{c}
\mp \sqrt{\frac{\left( j+1/2\right) \mp \left( M-1/2\right) }{2j+\left( 1\pm
1\right) }}Y_{j\pm 1/2,M-1/2}\left( \theta ,\varphi \right) \\
\sqrt{\frac{\left( j+1/2\right) \pm \left( M+1/2\right) }{2j+\left( 1\pm
1\right) }}Y_{j\pm 1/2,M+1/2}\left( \theta ,\varphi \right)%
\end{array}%
\right) ,
\end{equation}%
with the radial functions $f\left( r\right) $ and $g\left( r\right) $ are
given as:\qquad \qquad\

\begin{eqnarray}
\left(
\begin{array}{c}
f\left( r\right) \\
g\left( r\right)%
\end{array}%
\right) &=&\frac{\left( ma\right) ^{2}}{\nu }\sqrt{\frac{\left( E\varkappa
-m\nu \right) n!}{m\mu \left( \varkappa -\nu \right) \Gamma \left( n+2\nu
\right) }}e^{-\frac{1}{2}x}x^{\nu -1}\times  \notag \\
&&\times \left(
\begin{array}{c}
f_{1}xL_{n-1}^{2\nu +1}\left( x\right) +f_{2}L_{n}^{2\nu -1}\left( x\right)
\\
g_{1}xL_{n-1}^{2\nu +1}\left( x\right) +g_{2}L_{n}^{2\nu -1}\left( x\right)%
\end{array}%
\right) ,
\end{eqnarray}

where the relativistic energy levels are given by

\begin{equation}
E=E_{n,l}=\frac{m\left( n+\nu \right) }{\sqrt{\alpha ^{2}+\left( n+\nu
\right) ^{2}}}\qquad ,n=0,1,2...\text{\ \ }
\end{equation}%
and $L_{n}^{\alpha }\left( x\right) $ are the associated Laguerre
polynomials \cite{20}, with the following notations:

\begin{eqnarray*}
a &=&\frac{1}{m}\sqrt{m^{2}-E^{2}},\qquad \varkappa =\pm \left( j+\frac{1}{2}%
\right) ,\text{ \ \ \ \ \ }\nu =\sqrt{\varkappa ^{2}-\alpha ^{2}},\text{ \ \
}x=2\sqrt{m^{2}-E^{2}}\ , \\
f_{1} &=&\frac{a\alpha }{E\varkappa -m\nu },\qquad f_{2}=\varkappa -\nu
,\qquad g_{1}=\frac{a\left( \varkappa -\nu \right) }{E\varkappa -m\nu }%
,\qquad g_{2}=e^{2}=\alpha .
\end{eqnarray*}

\subsection{Noncommutative corrections of the energy}

Now to obtain the modification to the energy levels as a result of the terms
(\ref{a2}) due to the non-commutativity of space-space, we use perturbation
theory up to the first order. With respect the selection rules $\Delta l=0$
we have:

\begin{equation}
\Delta E_{n,l}=\Delta E_{n,l}^{\left( 1\right) }+\Delta E_{n,l}^{\left(
2\right) },
\end{equation}%
where:%
\begin{align}
\Delta E_{n,l}^{\left( 1\right) }& =-\frac{e^{2}}{2}\int_{0}^{4\pi }d\Omega
\int_{0}^{\infty }drr^{-1}[\psi _{njlM}^{\dagger }\left( r,\theta ,\varphi
\right) \left( \overrightarrow{\theta }\cdot \overrightarrow{L}\right) \psi
_{nj^{\prime }l^{\prime }M^{\prime }}\left( r,\theta ,\varphi \right) ]
\notag \\
& =-\frac{e^{2}}{2}\varrho _{n,l}^{\left( 1\right) }\Theta _{n,l,M,M^{\prime
}}^{\left( 1\right) },  \label{aa1} \\
\Delta E_{n,l}^{\left( 2\right) }& =\frac{e^{4}}{4}\int_{0}^{4\pi }d\Omega
\int_{0}^{\infty }drr^{-1}[\psi _{njlM}^{\dagger }\left( r,\theta ,\varphi
\right) [\overrightarrow{\alpha }\cdot (\overrightarrow{\theta }\times \frac{%
\overrightarrow{r}}{r})]\psi _{nj^{\prime }l^{\prime }M^{\prime }}\left(
r,\theta ,\varphi \right) ]  \notag \\
& =\frac{e^{4}}{4}\varrho _{n,l}^{\left( 2\right) }\Theta _{n,l,M,M^{\prime
}}^{\left( 2\right) },  \label{aa2}
\end{align}%
where%
\begin{align}
\varrho _{n,l}^{\left( 1\right) }& =\int_{0}^{+\infty }r^{-1}\left(
f^{2}+g^{2}\right) dr, \\
\varrho _{n,l}^{\left( 2\right) }& =\int_{0}^{+\infty }r^{-1}\left(
f^{2}-g^{2}\right) dr, \\
\Theta _{n,l,,M,M^{\prime }}^{\left( 1\right) }& =\int_{0}^{4\pi }d\Omega
\Omega _{jlM}^{\dagger }\left( \theta ,\varphi \right) (\overrightarrow{%
\theta }\cdot \overrightarrow{L})\Omega _{jlM^{\prime }}\left( \theta
,\varphi \right) , \\
\Theta _{n,l,,M,M^{\prime }}^{\left( 2\right) }& =\int_{0}^{4\pi }d\Omega
\Omega _{jlM}^{\dagger }\left( \theta ,\varphi \right) \overrightarrow{%
\sigma }\cdot (\overrightarrow{\theta }\times \frac{\overrightarrow{r}}{r}%
)\Omega _{jlM^{\prime }}\left( \theta ,\varphi \right) .
\end{align}

where the radial integrals are given by \cite{21}$:$%
\begin{align}
\varrho _{n,l}^{\left( 1\right) }& =\left( ma\right) ^{3}\left[ \frac{%
3E\varkappa \left( E\varkappa -m\right) -\left( \nu ^{2}-1\right) }{m^{2}\nu
\left( 4\nu ^{2}-1\right) \left( \nu ^{2}-1\right) }\right] , \\
\varrho _{n,l}^{\left( 2\right) }& =\frac{2\left( ma\right) ^{3}E}{m^{2}}%
\frac{m+2m\nu ^{2}-3E\varkappa }{\nu \left( 4\nu ^{2}-1\right) \left( \nu
^{2}-1\right) }.
\end{align}

The selection rules for the possible transitions between levels $\left(
Nl_{j}^{M}\rightarrow Nl_{j}^{M^{\prime }}\right) $ are $\Delta l=0$ and $%
\Delta M=0,\pm 1$, where $N=n+\left\vert \varkappa \right\vert $ describes
the principal quantum number. The $2P_{1/2}$ and $2P_{3/2}$ levels
correspond respectively to:
\begin{equation*}
\left( n=1,j=1/2,\varkappa =1,M=\pm 1/2\right)
\end{equation*}%
and
\begin{equation*}
\left( n=0,j=3/2,\varkappa =2,M=\pm 1/2,\pm 3/2\right) .
\end{equation*}%
The corresponding angular corrections are given by (we take $\theta
_{i}=\theta \delta _{i3}$):%
\begin{align}
\Theta _{2P_{1/2}}^{\left( 1\right) }& =\frac{2}{3}\theta \left(
\begin{array}{cc}
-1 & 0 \\
0 & 1%
\end{array}%
\right) ,\qquad \lambda _{2P_{1/2}}^{\left( 1\right) }=\pm \frac{2}{3}%
\left\vert \theta \right\vert ,  \label{cc1} \\
\Theta _{2P_{3/2}}^{\left( 1\right) }& =\frac{1}{3}\theta \left(
\begin{array}{cc}
-\Lambda & 0 \\
0 & \Lambda%
\end{array}%
\right) ,\qquad \Lambda =\left(
\begin{array}{cc}
3 & 0 \\
0 & 1%
\end{array}%
\right) ,\qquad \lambda _{2P_{3/2}}^{\left( 1\right) }=\pm \left\vert \theta
\right\vert ,\pm \frac{\left\vert \theta \right\vert }{3},  \label{cc2} \\
\Theta _{2P_{1/2}}^{\left( 2\right) }& =0,\qquad \Theta _{2P_{3/2}}^{\left(
2\right) }=0,  \label{cc3}
\end{align}%
where $\lambda _{2P_{1/2}}^{\left( 1\right) }$ and $\lambda
_{2P_{3/2}}^{\left( 1\right) }$ are respectively the eigenvalues of the
angular part.

From (\ref{aa1}), (\ref{aa2}), (\ref{cc1}) and (\ref{cc2}) we can write:%
\begin{align}
\Delta E_{2P_{1/2}}& =-\frac{e^{2}}{2}\varrho _{2P_{1/2}}^{\left( 1\right)
}\lambda _{2P_{1/2}}^{\left( 1\right) }=\mp 6.57668\times 10^{6}\left\vert
\theta \right\vert \text{ }\left( eV\right) ^{3},  \label{tt} \\
\Delta E_{2P_{3/2}}& =-\frac{e^{2}}{2}\varrho _{2P_{3/2}}^{\left( 1\right)
}\lambda _{2P_{3/2}}^{\left( 1\right) }=1.\,\allowbreak 578\times
10^{6}\left( \pm \left\vert \theta \right\vert ,\pm \frac{1}{3}\left\vert
\theta \right\vert \right) \left( eV\right) ^{3}.  \label{tt2}
\end{align}

According to Ref. \cite{22} the current theoretical accuracy on the $2P$
Lamb shift is about $0.08$ kHz. From the splitting (\ref{tt}), we get the
bound

\begin{equation*}
\theta \lesssim \left( 4GeV\right) ^{-2}
\end{equation*}%
and from the splitting (\ref{tt2}), we get the bound

\begin{equation*}
\theta \lesssim \left( 2GeV\right) ^{-2}\text{ or }\theta \lesssim \left(
1,2GeV\right) ^{-2}
\end{equation*}

It is worth mentioning that the second term of the perturbation in the
hamiltonian expression (\ref{a2}) does not  remove the degeneracy of the
energy levels because it is a non-diagonal matrix. However, for instance,
the non-vanishing matrix elements between $2S_{1/2}$ $\left(
n=1,j=1/2,\varkappa =-1,M=\pm 1/2\right) $ and $2P_{1/2}$ $\left(
n=1,j=1/2,\varkappa =1,M=\pm 1/2\right) $ states for the selection rules $%
\Delta l=1$ and $\Delta M=0,\pm 1$ give the possible transition:\ \ \ \
\begin{equation}
\langle 2P_{1/2}\left\vert \frac{e^{4}}{4}\overrightarrow{\theta }\cdot
\left( \overrightarrow{\alpha }\times \frac{\overrightarrow{r}}{r^{4}}%
\right) \right\vert 2S_{1/2}\rangle =\frac{e^{4}}{4}\Theta
_{2S_{1/2}\rightarrow 2P_{1/2}}\varrho _{2S_{1/2}\rightarrow 2P_{1/2}},
\label{45}
\end{equation}%
where%
\begin{equation}
\Theta _{2S_{1/2}\rightarrow 2P_{1/2}}=\frac{2}{3}\theta \left(
\begin{array}{cc}
1 & 0 \\
0 & -1%
\end{array}%
\right) ,  \label{46}
\end{equation}%
and%
\begin{equation}
\varrho _{2S_{1/2}\rightarrow 2P_{1/2}}=\frac{2\left( ma_{1}\right) ^{3}E_{1}%
}{m^{2}}\frac{m+2m\nu _{1}^{2}-3E_{1}\varkappa }{\nu _{1}\left( 4\nu
_{1}^{2}-1\right) \left( \nu _{1}^{2}-1\right) },  \label{47}
\end{equation}%
where

\begin{equation*}
\varkappa =1,\text{\ \ \ \ \ }\nu _{1}=\sqrt{1-e^{4}},\text{ \ \ \ \ }a_{1}=%
\frac{1}{m}\sqrt{m^{2}-E_{1}^{2}},\text{ \ \ \ \ \ }E_{1}=\frac{m}{\sqrt{%
1+\left( \frac{e^{2}}{1+\nu _{1}}\right) ^{2}}}.
\end{equation*}

From (\ref{45}), (\ref{46}) and (\ref{47}) there is an energy splitting of
the levels equal to

\begin{equation}
\Delta E_{2S_{1/2}\rightarrow 2P_{1/2}}=2\frac{e^{4}}{4}\frac{4\left(
ma\right) ^{3}E_{1}}{3m^{2}}\frac{m+2m\nu _{1}^{2}-3E_{1}}{\nu _{1}\left(
4\nu _{1}^{2}-1\right) \left( \nu _{1}^{2}-1\right) }\left\vert \theta
\right\vert \simeq \alpha \left\vert \Delta E_{2P_{1/2}}\right\vert
\end{equation}

This splitting is very similar to the anomalous Zeeman effect or Stark
effect at second order. Although the transition energy is small compared to
the Bohm shift $\Delta E_{2P_{1/2}}$. However is remains important in the
case of treatment of the hydrogen atom in the framework of noncommutative
QCD. This term is necessary to maintain the invariance of the modified Dirac
equation under the Seiberg-Witten maps.

\section{Non-relativistic limit of NC Dirac equation}

The non-relativistic limit of the noncommutative Dirac equation (\ref{aa})
corresponds to $\hat{\chi}\ll \hat{\varphi}$ \cite{16}, where, by restoring
the constants $c$ and $\hbar $, the wave function takes the new form
\begin{equation}
\hat{\psi}\left( t,r,\theta ,\varphi \right) =\hat{\psi}^{\prime }\left(
t,r,\theta ,\varphi \right) \exp \left( (-imc^{2}t/\hbar \right) ,
\end{equation}%
then, the non-relativistic form of the expression (\ref{aa})\ is given by the
following set of equations%
\begin{align}
\left( i\hbar \frac{\partial }{\partial t}-e\hat{\Phi}\right) \hat{\varphi}&
=c\overrightarrow{\sigma }\cdot \left( \overrightarrow{p}-\frac{e}{c}%
\overrightarrow{\hat{A}}\right) \hat{\chi}, \\
\left( i\hbar \frac{\partial }{\partial t}-e\hat{\Phi}+2mc^{2}\right) \hat{%
\chi}& =c\overrightarrow{\sigma }\cdot \left( \overrightarrow{p}-\frac{e}{c}%
\overrightarrow{\hat{A}}\right) \hat{\varphi},
\end{align}%
where%
\begin{equation}
\overrightarrow{\hat{A}}=\frac{e^{3}}{4\hbar c}\left( \overrightarrow{\theta
}\times \frac{\overrightarrow{r}}{r^{4}}\right) \qquad ,\qquad \hat{\Phi}%
=-\left( \frac{e}{r}+\frac{e}{2\hbar r^{3}}\overrightarrow{\theta }\cdot
\overrightarrow{L}\right) .
\end{equation}

If we consider the corrections up to the order of $1/c^{2}$, we can write
the Schrodinger equation of the bi-spinor $\hat{\varphi}$ as%
\begin{equation}
\hat{\varepsilon}\varphi _{nlM}^{sc}\left( t,r,\theta ,\varphi \right) =\hat{%
H}\varphi _{nlM}^{sc}\left( t,r,\theta ,\varphi \right) ,  \label{53}
\end{equation}%
where%
\begin{align}
\hat{H}& =\frac{1}{2m}\left( \overrightarrow{p}-\frac{e}{c}\overrightarrow{%
\hat{A}}\right) ^{2}+e\hat{\Phi}-\frac{p^{4}}{8m^{3}c^{2}}-\frac{e\hbar }{2mc%
}\overrightarrow{\sigma }\cdot \left( \overrightarrow{\nabla }\times
\overrightarrow{\hat{A}}\right) ,  \notag \\
& -\frac{e\hbar }{4m^{2}c^{2}}\overrightarrow{\sigma }\cdot \left(
\overrightarrow{\hat{E}}\times \overrightarrow{P}\right) -\frac{e\hbar ^{2}}{%
8m^{2}c^{2}}\overrightarrow{\nabla }\cdot \overrightarrow{\hat{E}}\qquad
,\qquad \overrightarrow{\hat{E}}=-\overrightarrow{\nabla }\hat{\Phi},
\label{54}
\end{align}%
and%
\begin{equation}
\varphi _{nlM}^{sc}\left( t,r,\theta ,\varphi \right) =R_{nl}\left( r\right)
\Omega _{j=l\pm \frac{1}{2},M}\left( \theta ,\varphi \right) ,
\end{equation}%
\begin{equation}
R_{nl}\left( r\right) =\frac{2}{n^{2}}\sqrt{\frac{\left( n-l-1\right) !}{%
a_{0}^{3}\left[ \left( n+l\right) !\right] ^{3}}}%
x^{l}e^{-x/2}L_{n-l-1}^{2l+1}\left( x\right) ,\ \ \ \ \ \ x=\frac{2r}{na_{0}}%
,\ \ \ \ \ a_{0}=\frac{\hbar ^{2}}{me^{2}},
\end{equation}%
\begin{equation}
\Omega _{j=l\pm \frac{1}{2},M}\left( \theta ,\varphi \right) =\left(
\begin{array}{c}
\pm \sqrt{\frac{l\pm M+\frac{1}{2}}{2l+1}}Y_{l,M-\frac{1}{2}}\left( \theta
,\varphi \right) \\
\sqrt{\frac{l\mp M+\frac{1}{2}}{2l+1}}Y_{l,M+\frac{1}{2}}\left( \theta
,\varphi \right)%
\end{array}%
\right) .
\end{equation}%
The energy corresponding to $\theta =0$ in the Schr\"{o}dinger equation (\ref%
{53}) is given by%
\begin{equation}
\varepsilon _{n}^{0}=-\left( \frac{e^{2}}{\hbar c}\right) ^{2}\frac{mc^{2}}{%
2n^{2}}\qquad ,\qquad n=1,2,3,...
\end{equation}

After a straightforward calculation, the equation (\ref{54}) takes the form:%
\begin{align}
\hat{H}& =\frac{p^{2}}{2m}-\frac{e^{2}}{r}-\frac{p^{4}}{8m^{3}c^{2}}-\frac{%
e^{4}}{\hbar mc^{2}r^{4}}\left( \overrightarrow{\theta }\cdot
\overrightarrow{L}\right) -\frac{e^{2}}{2\hbar r^{3}}\left( \overrightarrow{%
\theta }\cdot \overrightarrow{L}\right) +  \notag \\
& +\frac{e^{4}}{8mc^{2}r^{4}}\left[ \left( \vec{\sigma}\cdot \vec{\theta}%
\right) -\frac{4}{r^{2}}\left( \vec{\sigma}\cdot \overrightarrow{r}\right)
\left( \vec{\theta}\cdot \overrightarrow{r}\right) \right] +\frac{e^{2}\hbar
}{4m^{2}c^{2}r^{3}}\left[ \left( \overrightarrow{\sigma }\cdot
\overrightarrow{L}\right) \right. +  \notag \\
& \left. +\frac{3}{2\hbar r^{2}}\left[ \left( \overrightarrow{\theta }\cdot
\overrightarrow{L}\right) \left( \overrightarrow{\sigma }\cdot
\overrightarrow{L}\right) +\hbar \left( \overrightarrow{\sigma }\cdot
\overrightarrow{r}\right) .\left( \vec{\theta}\cdot \overrightarrow{p}%
\right) -\hbar \left( \overrightarrow{\sigma }\cdot \vec{\theta}\right)
.\left( \overrightarrow{r}\cdot \overrightarrow{p}\right) \right] \right] +
\notag \\
& +\frac{e^{2}\hbar ^{2}}{8m^{2}c^{2}}\left[ 4\pi \delta \left( r\right) -%
\frac{(\overrightarrow{\theta }\cdot \overrightarrow{L})p^{2}}{\hbar
^{3}r^{3}}+\frac{3}{\hbar ^{2}r^{5}}\left[ 2\hbar (\overrightarrow{\theta }%
\cdot \overrightarrow{L})\right. \right. -  \notag \\
& \left. \left. -\left( \overrightarrow{\theta }\cdot \overrightarrow{L}%
\right) \left( \overrightarrow{p}\cdot \vec{r}\right) -\vec{r}\cdot
\left( \overrightarrow{\theta }\cdot \overrightarrow{L}\right) \cdot
\overrightarrow{p}\right] \right] +O\left( \frac{1}{c^{3}}\right).
\label{59}
\end{align}
This hamiltonian is the non-relativistic limit of the one in eq.
(24), and contains new terms involving the parameter $\theta $ that
are similar to the ones of the ordinary hyperfine splitting: we can
say that the noncommutativity in this case plays the same role as
the spin interaction between the proton and the electron in the
presence of a magnetic field, which is responsible for the hyperfine
splitting.

Now to obtain the modification of energy levels as a result of the
non-commutative terms in eq. (\ref{59}), we use the first-order perturbation
theory. The expectation value of non-vanishing terms of the hamiltonian (\ref%
{59}) with respect to the solution in eq. (\ref{53}) are given by ( $\theta
_{i}=\theta \delta _{i3}$ and $l\neq 0$):

\begin{align}
\left\langle p^{4}\right\rangle & =-\frac{4m^{2}e^{4}}{n^{3}a_{0}^{2}}\left[
\frac{1}{l+1/2}-\frac{3}{4n}\right] ,  \notag \\
\left\langle \frac{\overrightarrow{\theta }\cdot \overrightarrow{L}}{r^{4}}%
\right\rangle & =\theta \hbar m_{j}\left( 1\mp \frac{1}{2l+1}\right)
\left\langle r^{-4}\right\rangle , \\
\left\langle \frac{\overrightarrow{\theta }\cdot \overrightarrow{L}}{r^{3}}%
\right\rangle & =\theta \hbar m_{j}\left( 1\mp \frac{1}{2l+1}\right)
\left\langle r^{-3}\right\rangle ,  \notag \\
\left\langle \frac{\vec{\sigma}\cdot \vec{\theta}}{r^{4}}\right\rangle &
=\pm \theta \frac{2m_{j}}{2l+1}\left\langle r^{-4}\right\rangle ,  \notag \\
\left\langle \frac{\left( \vec{\sigma}\cdot \overrightarrow{r}\right) (\vec{%
\theta}\cdot \overrightarrow{r}))}{r^{6}}\right\rangle & =\pm \theta \frac{%
2m_{j}}{\left( 2l+1\right) }\left[ \frac{\left( l+m_{j}+1/2\right) \left(
l-m_{j}+1/2\right) }{\left( 2l+1\right) ^{2}}+\right.  \label{60} \\
& \left. +\frac{\left( l+m_{j}+1/2\pm 1\right) \left( l-m_{j}+3/2\right) }{%
\left( 2\left( l\pm 1\right) +1\right) ^{2}}\right] \left\langle
r^{-4}\right\rangle ,  \notag \\
\left\langle \frac{\overrightarrow{\sigma }\cdot \overrightarrow{L}}{r^{3}}%
\right\rangle & =\hbar \left[ j\left( j+1\right) -l\left( l+1\right) -\frac{3%
}{4}\right] \left\langle r^{-3}\right\rangle ,  \notag \\
\left\langle \frac{\left( \overrightarrow{\theta }\cdot \overrightarrow{L}%
\right) \left( \overrightarrow{\sigma }\cdot \overrightarrow{L}\right) }{%
r^{5}}\right\rangle & =\theta \hbar ^{2}m_{j}\left( 1\mp \frac{1}{2l+1}%
\right) \times \left[ j\left( j+1\right) -l\left( l+1\right) -\frac{3}{4}%
\right] \left\langle r^{-5}\right\rangle ,  \notag \\
\pi \left\langle \delta \left( r\right) \right\rangle & =\frac{\left(
e^{2}m\right) ^{3}}{\hbar ^{6}n^{3}}\ \ \ \ \text{\ for}\ l=0\ \ \ \text{and}%
\ \ \ 0\ \ \text{for\ }\ l\neq 0,  \notag \\
\left\langle \frac{\left( \overrightarrow{\theta }\cdot \overrightarrow{L}%
\right) p^{2}}{r^{3}}\right\rangle & =2\theta me^{2}\hbar m_{j}\left( 1\mp
\frac{1}{2l+1}\right) \times \left[ \frac{1}{2a_{0}n^{2}}\left\langle
r^{-3}\right\rangle +\left\langle r^{-4}\right\rangle \right] ,  \notag \\
\left\langle \frac{\overrightarrow{\theta }\cdot \overrightarrow{L}}{r^{5}}%
\right\rangle & =\theta \hbar m_{j}\left( 1\mp \frac{1}{2l+1}\right)
\left\langle r^{-5}\right\rangle .  \notag
\end{align}%
where we have $\overrightarrow{S}=\frac{\hbar }{2}\overrightarrow{\sigma },$
$\left\langle S_{z}\right\rangle =\pm \frac{\hbar m_{j}}{2l+1}$ , $%
\left\langle L_{z}\right\rangle =\left\langle
J_{z}-S_{z}\right\rangle =\hbar m_{j}\left( 1\mp
\frac{1}{2l+1}\right) $ and the positive and negative sings
correspond to $j=l+1/2$ and $j=l-1/2$ respectively. In the above
$a_0$ is the Bohr radius.

Finally the first-order energy correction is
\begin{equation}
\Delta\varepsilon\left( l\neq0\right) =\Delta\varepsilon_{0}+\Delta
\varepsilon_{\theta}.
\end{equation}
The first term $\Delta\varepsilon_{0}$ represents the ordinary
fine-structure correction and is given by:

\begin{equation}
\Delta \varepsilon _{0}=\frac{e^{4}}{2mc^{2}}\frac{1}{n^{3}a_{0}^{2}}\left[
\frac{1}{l+1/2}-\frac{3}{4n}\right] +\frac{e^{2}\hbar ^{2}}{4m^{2}c^{2}}%
\left[ j\left( j+1\right) -l\left( l+1\right) -\frac{3}{4}\right]
\left\langle r^{-3}\right\rangle .
\end{equation}%
The last term $\Delta \varepsilon _{\theta }$ is very similar to that of the
hyperfine structure correction, where $\theta $ is now replacing spin and
magnetic field, and is given by:

\begin{align}
\Delta \varepsilon _{\theta }& =\frac{\theta }{2}e^{2}m_{j}\left\{ \left( -1+%
\frac{e^{4}}{4\hbar ^{2}c^{2}}\frac{1}{n^{2}}\right) \left( 1\mp \frac{1}{%
2l+1}\right) \times \left\langle r^{-3}\right\rangle -\right.  \notag \\
& -\frac{e^{2}}{2mc^{2}}\left[ \left( 5\pm \frac{6}{2l+1}\right) \pm \frac{4%
}{\left( 2l+1\right) }\left[ \frac{\left( l+m_{j}+1/2\right) \left(
l-m_{j}+1/2\right) }{\left( 2l+1\right) ^{2}}+\right. \right.  \notag \\
& \left. \left. +\frac{\left( l+m_{j}+1/2\pm 1\right) \left(
l-m_{j}+3/2\right) }{\left( 2\left( l\pm 1\right) +1\right) ^{2}}\right] %
\right] \times \left\langle r^{-4}\right\rangle +  \notag \\
& +\left. \frac{3\hbar ^{2}}{4m^{2}c^{2}}\left( 1\mp \frac{1}{2l+1}\right)
\left( j\left( j+1\right) -l\left( l+1\right) +\frac{5}{4}\right) \times
\left\langle r^{-5}\right\rangle \right\} .
\end{align}%
where%
\begin{align}
\left\langle r^{-3}\right\rangle & =\frac{1}{a_{0}^{2}n^{3}}\frac{1}{l\left(
l+1/2\right) \left( l+1\right) }, \\
\left\langle r^{-4}\right\rangle & =\frac{2}{a_{0}^{4}n^{3}}\frac{1}{\left(
2l+3\right) \left( 2l-1\right) \left( l+1/2\right) }\left[ -\frac{1}{n^{2}}+%
\frac{3}{l\left( l+1\right) }\right] , \\
\left\langle r^{-5}\right\rangle & =\frac{1}{3a_{0}^{5}n^{3}}\frac{1}{\left(
l+2\right) \left( l-1\right) \left( l+1/2\right) }  \notag \\
& \qquad \times \left\{ -\frac{2}{n^{2}}\frac{1}{l\left( l+1\right) }+\frac{5%
}{\left( 2l+3\right) \left( l-1/2\right) }\left[ -\frac{1}{n^{2}}+\frac{3}{%
l\left( l+1\right) }\right] \right\} ,
\end{align}

This result shows that, in the non-commutative non-relativistic theory, the
degeneracy is completely removed and describes the correction of the fine
structure of the spectrum, and corresponds to the hyperfine splitting. Thus
by comparing to the data one can get an experimental bound on the value of $%
\theta $.

For the case $l=0$ all terms in eq. (\ref{60}) vanish except the fourth and
fifth ones. These terms give a divergence and hence we use the $\Lambda _{%
\text{QCD}}$ ($\sim 200$MeV) cutoff (see ref \cite{23}) and obtain the
result:

\begin{equation}
\left\langle \left( (\vec{\sigma}\cdot \vec{\theta})/r^{4}-4\left( \vec{%
\sigma}\cdot \overrightarrow{r}\right) (\vec{\theta}\cdot \overrightarrow{r}%
)/r^{6}\right) \right\rangle _{\text{1S}}=\frac{4\theta }{3}\alpha
^{3}m^{3}\Lambda _{\text{QCD}}.  \label{67}
\end{equation}

From equation $\left( 65\right) $ we obtain the modified energy level in
noncommutative space-space in the non-relativistic limit for the state 1S:

\begin{equation}
\Delta \varepsilon _{\theta }=\frac{\theta }{6}\alpha ^{5}m^{2}\Lambda _{%
\text{QCD}}
\end{equation}

According to Ref. \cite{22} the current theoretical accuracy on the 1S Lamb
shift is about $14$ kHz. From the splitting (\ref{67}), the bound is given by

\begin{equation}
\theta \lesssim \left( 5.6GeV\right) ^{-2}
\end{equation}

This value is better than the limit obtained in \cite{14, 23} and it
justifies our expansion of the Hamiltonian in\ eq (\ref{a2}) .

\section{Conclusions}

In this work we proposed an invariant noncommutative action for a
Dirac particle under the generalised infinitesimal gauge
transformations. Using the Seiberg-Witten maps and the Moyal
product, we generalised the equation of motion with a noncommutative
space-space and derived the modified Dirac equation for a Coulomb
potential to the first order of $\theta $. By perturbation-theory
methods in first order, we derived the noncommutative corrections of
the energy. In addition to the hamiltonian given in \cite{10} where
the authors have used the noncommutative Bopp-shift, another term
appears in the Hamiltonian which is similar to the interaction term
describing charged particles in a non-zero magnetic field. This
non-diagonal term is a vectorial potential due to the invariance of
the modified Dirac equation under the Seiberg-Witten maps. In this
case the degeneracy of energy-level states is removed and the
lamb-shift is induced. The bound we found on $\theta$ has the same
order of magnitude obtained in ref. \cite{10}.

In the non-relativistic limit, we have obtained a general modified
form of the hamiltonian of the hydrogen atom with new terms
involving the $\theta$ parameter. This expression is similar to the
hyperfine structure one. The expression of the hamiltonian (in the
non-relativistic limit) which describes the hyperfine correction in
the hydrogen atom imply that the noncommutativity plays the role of
the magnetic field (Zeeman effect) and the role of the spin (of the
proton or nucleon). Then the interaction electron-nucleon is
equivalent to an electron in a noncommutative space-space. In this
case, the degeneracy of the energy-level states is completely
removed and the bound on $\theta$ was derived.

\end{document}